\documentstyle[preprint,epsfig,aps,floats]{revtex}
\draft  
\begin{document}
%%\twocolumn[\hsize\textwidth\columnwidth\hsize\csname
%%           @twocolumnfalse\endcsname

\title{Quantum gravity effects near the  
null black hole singularity}
\author{A. Bonanno}
\address{Istituto di Astronomia, Universit\`a di Catania\\
Viale Andrea Doria 6, 95125 Catania, Italy\\
INFN Sezione di Catania, Corso Italia 57, 95100 Catania, Italy\\}
\author{M. Reuter}
\address{Institut f\"ur Physik, Universit\"at Mainz\\
Staudingerweg 7, 55099 Mainz, Germany\\}

\maketitle
\begin{abstract}
\widetext
The structure of the Cauchy Horizon singularity of a black hole 
formed in a generic collapse is studied by means of a 
renormalization group equation for quantum gravity. 
It is shown that during the early evolution of the Cauchy Horizon 
the increase of the mass function is damped when quantum fluctuations 
of the metric are taken into account.
\end{abstract}

\vskip 2pc
\narrowtext
%%]
\def\be{\begin{equation}}
\def\ee{\end{equation}}
\def\bea{\begin{eqnarray}}
\def\eea{\end{eqnarray}}
\def\RN{Reis\-sner-Nord\-str\"{o}m }

\section{introduction}
Recently much progress has been made in understanding the formation
of singularities in realistic black holes.
After the seminal work by Poisson and Israel
\cite{pi}, the outcome of several investigations with spherical models 
(see \cite{haifa} for a general overview) was 
that the spacetime develops a null scalar singularity at the Cauchy 
Horizon (CH) whose subsequent evolution eventually 
stops at the final spacelike singularity at $r=0$ \cite{abo}.

In particular the Petrov type D component $\Psi_2$ of the
Weyl curvature diverges exponentially with advanced time 
at this lightlike hypersurface, although 
the ``measured" tidal distortion is bounded. 
The metric tensor is regular in 
a suitable local chart adapted to the inner horizon and the metric
perturbations are small. This scenario is likely to 
be essentially the same in more general contexts than 
spherical symmetry \cite{ori}, but the structure of the singularity 
is more complicated then since the square of the Weyl 
tensor $C_{\mu\nu\tau\lambda}C^{\mu\nu\tau\lambda}$
is dominated by the radiative component $\Psi_0\Psi_4$ of a Petrov 
type N curvature \cite{droz}.

It is an interesting question  
whether quantum effects can modify the classical evolution 
of the fields in a significant way. 
As far as the classical evolution is concerned, causality does
not permit our ignorance about the correct form of the dynamics
in the inner, Planckian curvature regions of the
interior to infect the description of the overlaying layers in 
terms of classical general relativity. The radial coordinate $r$ 
is in fact timelike in the interior of a spherical hole.

This picture changes in quantum field theory because in loop
calculations even states localized outside the light-cone have an impact 
on the value of the renormalized quantities.
One can then imagine that the metric fluctuations near the 
inner horizon modify the infrared 
region where the Weyl curvature is still 
growing but it has not yet reached Planckian levels. 
In particular it is interesting to see if the presence of 
some ``self-regulator" mechanism could prevent the local 
curvature from diverging at the CH. 
An indication has been given in \cite{werlett} where it has been 
noticed that the classical divergence of the mass function 
in an evaporating black hole (BH) can be damped out by 
the contribution of the blueshifted influx 
of the Hawking radiation at late advanced times. 
More complete investigations in four dimensions \cite{israel} have 
been performed in the semiclassical approximation by considering a 
massless minimally coupled scalar field,
but they were inconclusive 
about the ``sign" of the quantum correction, {\em i.e.} about 
whether it would lead to a stronger or to a weaker divergence. 

The objective of the present investigation is to show that
quantum fluctuations of the gravitational field indeed weaken the strength of 
the singularity at the inner horizon. This result is obtained by studying the running of the
Newton constant at large momenta by means of  the non-perturbative renormalization-group 
equation \cite{corfu,rw,aboza} which governs the scale dependence of the effective average 
action $\Gamma_k$ for gravity \cite{reuter}. $\Gamma_k$ is a Wilson-type effective action with a
built-in infrared (IR) cut-off at the mass scale $k$. The functional
$\Gamma_k$ is obtained by integrating out the quantum fluctuations
with momenta between a fixed ultraviolet (UV) cutoff $k_{\rm UV}$ and
the variable IR cutoff $k$. In this framework, a renormalizable theory 
with the classical action $S$ is quantized by solving the flow equation
subject to the initial condition $\Gamma_{k_{\rm UV}}=S$ and letting
then $k_{\rm UV}\rightarrow \infty$, $k\rightarrow 0$ (after suitable
renormalization). 

What makes the effective average action an ideal
tool for studying quantum gravity is the fact that this method can 
also be used in order to renormalization group-evolve (coarse grain)
the actions of non-renormalizable effective field theories. In this
case one assumes that there is some fundamental theory which has been 
``partially quantised'', {\em i.e.} its quantum fluctuations with momenta
from infinity down to a fixed scale $k_{\rm UV}$ have been integrated 
out already. This leads to an effective action $S_{\rm eff}$ 
which, when evaluated in tree approximation, 
correctly describes all phenomena with 
typical momenta of the order $k_{\rm UV}$. If we are interested in processes
at smaller momenta $k<k_{\rm UV}$ we can construct a new effective action,
appropriate for the lower scale, by setting 
$\Gamma_{k_{\rm UV}}=S_{\rm eff}$
and solving the flow equation for $\Gamma_{k}$ with this initial condition.
It is clear that for effective theories the limit 
$k_{\rm UV} \rightarrow \infty$ should not be performed; hence the 
non-renormalizability of a theory does not pose any problems 
in this context.

Quite generally, the effective action $\Gamma$~or the
average action $\Gamma_k$~encapsulates {\it all} physical effects of a
given theory.  Once we have identified its leading terms for a given range
of momenta, no other quantum corrections beyond those which are already
contained in the running coupling constants parametrising the approximate
form of $\Gamma_k$~ need to be taken into account. In the case at hand, a
truncated derivative expansion in powers of the curvature tensor and its
covariant derivatives is a sensible approximation as long as $k<<m_{\rm p}$
since the terms omitted are suppressed by higher powers of $k/m_{\rm p}$. 
($m_{\rm p}$ denotes the Planck mass.) Within
this approximation, the most relevant term in the action is the Einstein-Hilbert term
since it has the smallest canonical dimension. As a consequence, the most
important effects of quantum gravity are encoded in the running of the
associated coupling, i.e. Newton's constant. Thus, when we increase $k$ from
the classical ({\it i.e}, IR) domain to larger values in order to explore
gravity at smaller distances, the first sign of a non-classical behaviour
is a changing value of Newton's constant. In the present paper
we investigate the regime of momenta where, on the one hand, the first quantum gravitational 
effects appear already while on the other hand higher order invariants 
($R^2$ terms, etc.) are not yet important.

In the following we shall consider Einstein gravity as an effective field
theory and we identify the standard Einstein-Hilbert action with the average action
$\Gamma_{k_{\rm obs}}$. Here ${k_{\rm obs}}$ is some typical ``observational scale'' 
at which the classical tests of general relativity have confirmed the Einstein-Hilbert action.
We assume that also for $k> k_{\rm obs}$, {\em i.e.}
at higher energies, $\Gamma_{k}$ is well approximated by an action of 
the Einstein-Hilbert form as long as $k$ is not too close to the Planck scale.
The two parameters in this action, Newton's constant and the cosmological
constant, will depend on $k$, however, and the flow equation will tell us 
how the running Newton constant $G(k)$ and the running cosmological constant
$\Lambda(k)$ depend on the cutoff. Their experimentally observed values
are $G(k_{\rm obs})=G_{\rm obs}$ 
and $\Lambda(k_{\rm obs})=\Lambda_{\rm obs} \simeq 0$.
The Newton constant defines the Planck mass according to 
$m_{\rm p}=1/\sqrt{32\pi G_{\rm obs}}$. 

We fix a scale $k_{\rm UV}\gg k_{\rm obs}$ 
in such a way that it is still sufficiently
below the Planck scale so that $\Gamma_{k}$ is not yet very different 
from the Einstein-Hilbert action, but is already large enough for quantum
gravitational effects to play a role. We start the renormalization group
evolution at the scale $k_{\rm UV}$ with bare parameters 
$G(k_{\rm UV})=\bar{G}$ and $\Lambda(k_{\rm UV})=\bar{\Lambda}$ 
which should be thought of as functions of $G_{\rm obs}$
and $\Lambda_{\rm obs}$. We shall use our result for the function
$G(k), k\in [k_{\rm obs},k_{\rm UV}]$, in order to study the impact of the
scale dependent Newton constant on the mass-inflation scenario.

In a sense, we shall ``renormalization group improve'' the classical 
metric describing the late-time behavior of the spacetime near the CH.
%%%
%%% REVISION: an example from QED
%%%
Our method is similar to the following renormalization-group based
derivation of the Uehling correction to the Coulomb potential in
massless QED \cite{ue}. One starts from the classical potential 
energy $V_{\rm cl}(r) = e^2/4\pi r$ and replaces $e^2$ by the running
gauge coupling in the one-loop approximation: 
\be\label{uel}
e^2(k)=e^2(k_0)[1-b\ln(k/k_0)]^{-1},\;\;\;\;\; b\equiv e^2(k_0)/6\pi^2.
\ee
Hereby one may identify the renormalization point $k$ with the inverse of the 
distance $r$ because in the massless theory this is the only relevant scale.
The result of this substitution reads
\be\label{rea}
V(r)=-e^2(r_0^{-1})[1+b\ln(r_0/r) + O(e^4)]/4\pi r
\ee
where the IR reference scale $r_0\equiv 1/\mu_0$ has to be kept finite in the
massless theory. Note that eq.(\ref{uel}) is the correct (one-loop, massless)
Uehling potential which is usually derived by standard perturbative methods
\cite{ue}. Obviously the position dependent renormalization group improvement
$e^2\rightarrow e^2(k)$, $k\propto 1/r$ encapsulates the most important effects
which the quantum fluctuations have on the electric field produced by a point
charge. We argue that analogous substitution $G\rightarrow G(k)$ with an appropriate
$k=k(x^{\mu})$ yields the leading modification of the spacetime metric. 

\section{quantum gravity effects behind the inner potential barrier}
Considering a spherically symmetric, charged BH 
the metric can be conveniently 
expressed using the coordinates $x^a\;(a,b=0,1)$ on the radial two-spaces 
$(\theta,\phi)$ = {\it const}  
and the function $r(x^a)$ that measures the area of those two-spheres
whose line element is $r^2 d\Omega^2$. The metric element is then
$ ds^2= g_{ab}dx^a dx^b+r^2d\Omega^2.$ 
By defining the scalar fields $f(x^a),m(x^a)$ and $-2\kappa(x^a)=
\partial f/\partial r$  through 
$f=1-{2 G_{\rm obs}m/r}+{G_{\rm obs} e^2/ r^2}$
the Einstein equations reduce to the two-dimensionally covariant equations
\bea\label{einstein}
&&r_{;ab}+\kappa g_{ab}=-4\pi G_{\rm obs} r 
(T_{ab}-g_{ab}T)\nonumber\\[2 mm]
&&R-2\partial_r \kappa = 8\pi G_{\rm obs}(T-2P)
\eea
where the static electro-magnetic field is generated by a charge
of strength $e$ and $T_{ab}$ is the stress-energy tensor of the matter field 
whose two-dimensional trace is $T$ and tangential pressure is $P$.
From the conservation laws one finds the following 
two-dimensional wave equation for the mass function
\bea\label{wave}
&&\Box m=-16\pi^2r^3G_{\rm obs}T_{ab} T^{ab} 
+8\pi G_{\rm obs} f(P-T)\nonumber\\ 
&&+4\pi r^2 G_{\rm obs}\kappa T-4\pi r^2G_{\rm obs}r_{,a}T^{,a}.
\eea 
This latter equation is the key to understanding 
the phenomenon of the mass-inflation. 
The late time behavior of the external gravitational field produced during
the collapse of a star is that of a (Kerr-Newman, in general) 
black hole of external mass $m_0$ perturbed 
by a tail of gravitational waves 
whose flux decays as $\sim v^{-p}$ with $p = 4(l+1)$ for a 
multi-pole of order $l$.
As a consequence of the boundary conditions set at the
event horizon, the $T_{ab}T^{ab}$ interaction term 
between the influx and out-flux of gravitational waves 
scattered from the
inner potential barrier triggers a divergent source term
for the local mass function $m(u,v)$.
The outflow can be modelled as a radial stream of light-like 
material particles because of the infinite blue-shift near the Cauchy
Horizon. It is possible to show \cite{pi} that near the CH
\be\label{www}  
m(v,r)\sim v^{-p}e^{\kappa_0 v} \;\;\; (v\rightarrow \infty)
\ee
where $v$ is the standard advanced time Eddington-Finkelstein
coordinate. ($\kappa_0$ denotes the surface gravity of the
\RN static black hole that characterises the external field
configuration.) 

It must be observed that in Eq.(\ref{wave}) 
the strength of the gravitational interactions between 
out-flux and influx is proportional to the Newton constant. 
Hence small changes in $G$ due to renormalization effects 
are then exponentially amplified by the mass
function like in a magnifying lens!
In particular if gravity is asymptotically free 
the classical divergence of the mass function can be weakened by 
the decreasing of the Newton constant at small distances.

%%
%% REVISION insert figure
%%
\begin{figure}[h]
\hbox to\hsize{\hss\epsfxsize=7cm\epsfbox{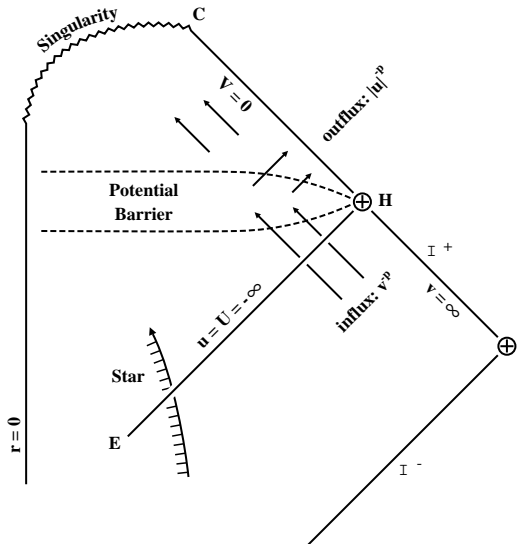}\hss}
\caption{Penrose conformal diagram of a collapsing star. Note that the point H
is not part of the manifold, but a singular point of this mapping.
\label{corn}}
\end{figure}
In order to discuss this phenomenon in the mass-inflation scenario we consider the model 
analysed in \cite{abo} for the scalar field collapse although our result 
should not depend on this particular framework. We are interested in the
asymptotic portion of the spacetime at late retarded times 
(the ``corner" region near the point H in Fig.(\ref{corn})) 
before the strong focusing region where $r \rightarrow 0$. 
The null Kruskal coordinates $U,V$ are thus introduced, 
being 
\be
\kappa_0 U=-\exp(\kappa_0 u),\;\;\;\;\;\; \kappa_0 V=-\exp(\kappa_0 v)
\ee 
were $(u,v)$ are the retarded and advanced time coordinates.
In a neighbourood of $(U=-\infty, V=0)$ an approximate analytical
solution of the Einstein equations and the
wave equation for a massless minimally coupled 
scalar field $\Phi$ can be found \cite{abo}. 
The explicit asymptotic expression for the metric is
\bea\label{metric}
&&ds^2 = -2 {r_0 \over r} dU dV + r^2 d\Omega^2\\
&&r^2  = r_0^2 -2G_{\rm obs}[A(U)+B(V)]
\eea
where $r_0$ is the location of the CH in the static 
BH spacetime configuration.
The dimensionless functions $A(U)$ and $B(V)$ are regular at the CH,
$A(-\infty)=B(0)=0$, but $\dot{B}$ diverges like $1/V(-\ln (-\kappa_0 V))^{(p+2)}$ 
as $V\rightarrow 0^{-}$ while $A$ is positive definite and $\dot{A}$ is bounded.
Even though the metric coefficients and the scalar field $\Phi$ are both regular 
at the CH, the mass function is divergent 
for $V\rightarrow 0^{-}$ being
\be\label{mass2}
m(U,V) \simeq  {G_{\rm obs} \over r_0}\dot{A}\dot{B}
\ee

We consider the evolution of the
above geometry in the mass-inflating regime starting from
a value of the coordinate $V=V_{\rm IR}$ for which the mass function 
is already exponentially growing $m(U,V)/m_0\gg 1$  
(we assume $r_0^{-1}\sim m_0 \gg 1$ in Planck units)
but the curvature has not yet reached Planckian values.

At this point we need an explicit expression for the running Newton
constant. We use the result obtained in \cite{reuter} where, for pure
gravity, the evolution of $\Gamma_k$ has been obtained in the 
``Einstein-Hilbert approximation'' 
where only the $\sqrt{-g}$ and $\sqrt{-g}R$ operators are 
considered in the renormalization group flow. 
%%
%% REVISION : new eq.
%%
This amounts to truncating the space of all the actions to those
of the form 
\be\label{ansa}
\Gamma_k[g,\bar{g}]=(16\pi G(k))^{-1}
\int d^4x\sqrt{g}\{ -R(g)+2\Lambda(k)\}+S_{\rm gf}[g,\bar{g}]
\ee
where $S_{\rm gf}$ is the classical background gauge fixing term.
For this truncation the flow equation reads
\bea\label{evo1}
&&k \partial_k \Gamma_k[g,\bar{g}] =
{1\over 2}{\rm Tr}
\Big [(\Gamma^{(2)}_k[g,\bar{g}]+{\cal R}^{\rm grav}_{k}[\bar{g}])^{-1}
k\partial_k {\cal R}^{\rm grav}_{k}[\bar{g}]\Big]\nonumber\\[2mm]
&&-{\rm Tr}\Big[(-{\cal M}[g,\bar{g}]+{\cal R}^{\rm gh}_{k}[\bar{g}])^{-1}
k\partial_k {\cal R}^{\rm gh}_k[\bar{g}] \Big ]
\eea
where $\bar{g}_{\mu\nu}$, $\Gamma_k^{(2)}$ and ${\cal M}$ denote the
background metric, the Hessian of $\Gamma_k$ with respect to the 
``ordinary'' metric  argument $g_{\mu\nu}$, and the Faddeev-Popov 
ghost operator, respectively. The operators ${\cal R}_k^{\rm grav}$ 
and ${\cal R}^{\rm gh}_k$ are the IR cutoffs in the graviton
and the ghost sector, respectively. They are defined in terms of 
an arbitrary smooth function ${\cal R}_k(p^2)$
(interpolating between zero for $p^2\rightarrow\infty$ 
and a constant $\propto k^2$ at $p^2=0$) 
by replacing $p^2$ with the graviton and ghost kinetic
operator, respectively. Inside loops, they suppress the contribution
from modes with covariant momenta $p<k$.

Upon projecting the renormalization group flow on the two dimensional
space spanned by the operators $\sqrt{-g}$ and $\sqrt{-g}R$
the functional flow equation becomes two ordinary differential
equations for $G(k)$ and $\Lambda(k)$. The equation for the 
scale-derivative of the running dimensionless Newton constant
$g(k)=k^2G(k)$ is found to be
\be\label{flw}
k\partial_k g(k) = [2+\eta(k)]g(k) 
\ee
where $\eta(k)\equiv gB_1/(1-gB_2)$ is an anomalous dimension
involving two known functions \cite{reuter} of the cosmological
constant, $B_1$ and $B_2$, which depend on the choice for 
${\cal R}_k(p^2)$.
Contrary to the running of the dimensionless gauge coupling 
$e(k)$ in QED, the beta-function describing the running of the 
Newton constant is not universal. It is scheme dependent even in the
lowest order of the loop expansion. In our framework this is reflected
by the ${\cal R}_k$-dependence of $\eta$. To lowest order of an expansion
in powers of $k/m_{\rm p}$ one may ignore the impact of the running cosmological
constant on $\eta(k)$ and set $\Lambda(k)\simeq 0$. Thus, returning 
to physical units and retaining only the leading term of the 
$k/m_{\rm p}$-expansion the solution to eq.(\ref{flw}) reads
\be\label{renorm}
G(k)=G_{\rm obs}[1-\omega G_{\rm obs}k^2
+O(k^4/m_{\rm p}^4)].
\ee
For pure gravity one obtains $\omega=\omega_G\equiv-B_1(\Lambda=0)/2>0$
which assumes the numerical value $\omega_G=4(1-\pi/144)/\pi$ for
a standard exponential cutoff \cite{reuter}. While $\omega$ 
depends on the shape of ${\cal R}_k$, it can be shown that $\omega$
is positive for any choice of this function. Consequently pure gravity 
is ``antiscreening'': Newton's constant decreases as $k$ increases,
{\em i.e.} it is large in the IR and becomes smaller in the UV.

Eq.(\ref{renorm}) is believed to be reliable as long as 
$k_{\rm UV}$ is still below the Planck mass. 
%%
%% REVISION : new comments on the domain of validity of the approximation.
%%            new references added
%%
If $k<k_{\rm UV}<<m_{\rm p}$ the effect, in the renormalised system, 
of higher curvature invariants such as $R^2$, $R_{\mu\nu}R^{\mu\nu}$ or $R^3$ 
which were omitted from the ansatz (\ref{ansa}) is indeed small. 
In fact, those invariants have been classified accordingly to their  
anomalous scaling  dimension which characterise the linearised renormalization group flow 
near the Gaussian fixed point \cite{aboza}. The result is that the flow in the UV
region is determined only by the ``relevant'' operators $\sqrt{g}$ and
$\sqrt{g} R$, and that any other invariant with a higher canonical dimension is suppressed 
by additional powers of $k/m_{\rm p}$ \cite{aboza}.
We shall then assume that 
\be\label{domain}
k<k_{\rm UV}=m_{\rm p}/a
\ee
with $a$ a fixed number well above unity.
This defines the domain of validity of our approximation. 

It is straightforward to include matter fields. 
In our model it might appear natural to keep the electro-magnetic
field classical but quantise the full scalar field. The only
effect on the running of $G$ is to shift the parameter $\omega$.
Using the same cut-off as above, one finds \cite{granda}
$\omega=\omega_{GS}=4/\pi-3\pi/72$ which, again, is positive
and leads to the same qualitative features as pure gravity.

The running of $G$ has dramatic 
consequences for the mass-inflation scenario.
The leading quantum correction of the metric is obtained by replacing
$G_{\rm obs}$ in eq.(\ref{mass2}) for the mass function by the
running Newton constant $G(k)$ with an appropriately 
chosen scale $k$.
Since $G(k)<G_{\rm obs}$ for any value of $k>k_{\rm obs}$ 
we conclude that {\it the quantum corrections tend to damp the increase of the 
mass function}. This qualitative conclusion is independent of the precise
definition of the cut-off $k$. It is a rather robust results therefore.

\section{an improved model}
%%
%% REVISION: we have tried to clarify several points
%%
We are now going to implement this mechanism in an iterative 
calculation, where the zeroth-order solution for the metric is substituted into
the running of $G$ in order to calculate the first-order correction of the metric.
For the sake of simplicity let us now consider the 
simpler case of the cross-flow model discussed in 
\cite{abo}. 

The first question to be answered is what is the analogue of the identification
$k\propto 1/r$ which we used in QED. We are looking for a $x^{\mu}$-dependent
cut-off $k=k(x^{\mu})$ which respects general coordinate
invariance and which measure the typical mass scale set by the curvature 
of spacetime. Since in the case at hand the metric is spherically symmetric,
the natural candidate for the cut-off is the ``coulombian'' component 
of the Weyl curvature 
\cite{israel}: $k^2 \propto |\Psi_2| = G_{obs}m(U,V)/r_{0}^3$. More precisely we
use the classical metric (zeroth-order approximation) to define the
position dependent IR cutoff by 
\be\label{cih}
k^2(V)=\max_U \{ b^2|\Psi_2| \}=
\max_U \{ b^2G_{\rm obs}m(U,V)/r^3_0 \}
\ee 
with $b$ another fixed number much larger than unity and the maximum is performed over 
the region near $U\rightarrow -\infty$. Here we are invoking a kind of adiabatic approximation 
where the use of a position dependent cutoff is justified because the mass function 
$m(U,V)$ is almost constant on the length scales at which the eigenmodes integrated 
out are varying. (A similar approximation has already been used in \cite{israel} in 
a semi-classical calculation.) From $k(V)$ one obtains a running Newton constant as 
a function of the $V$ coordinate
\be\label{guv}
G(V)=G_{\rm obs}
\Big[1-{\cal A}\dot{B}(V)\Big]
\ee
where 
\be
{\cal A}=\omega b^2 \max_U \{ G_{\rm obs}^3 
{\dot A}(U)/r^4_0 \}.
\ee
It is now possible to evolve the classical geometry in 
eq.(\ref{metric}) by considering the running Newton's constant in 
the Einstein equations. Within our approximation, the
improvement amounts to replacing 
\be
G_{\rm obs}T_{ab}=G_{\rm obs}T_{ab}^{\rm in}(V)+
G_{\rm obs}T_{ab}^{\rm out}(U)
\ee 
with 
\be  
T_{ab}^{\rm imp}=
G(V)T_{ab}^{\rm in}(V)+
G_{\rm obs}T_{ab}^{\rm out}(U).
\ee 
This modified energy-momentum tensor is then covariantly conserved 
since it satisfies $(T_{ab}^{\rm imp} r^2)^{;b}=0$.

From the Bianchi identities one obtains the following wave equation for the mass function
\bea\label{wave2}
&&\Box(Gm) = -16\pi^2r^3G^2T_{ab}T^{ab}+
e^2\Big ({G_{,a}\over 2 r}\Big )^{;a}.
\eea
The general solution is uniquely determined once the value of the fields
along the characteristic $U=U_{\rm IR}$ and $V=V_{\rm IR}$ 
are given. Asymptotically the improved metric is still of
the form (\ref{metric}) but eq.(8) is now replaced by
\be
r^2=r^2_0-2G(V)B(V)-2G_{\rm obs}A(U).
\ee
By noticing that $G$ approaches its bare value very rapidly
one finds that the leading term (as $V\rightarrow 0^{-}$) on the right hand side of 
eq.(\ref{wave2}) is now given by the classically divergent 
$T_{ab}T^{ab}$  contribution. Thus, after the inner potential barrier, 
one finds the solution
\be
m(U,V) \simeq {G_{\rm obs}
\over r_0}(1-{\cal A}\dot{B})\dot{A}\dot{B}-
{e^2 G_{\rm obs}\over 4 r_0^3}{\cal A}\dot{B}A
\ee
which replaces the classical expression (\ref{mass2}). 
%%
%% REVISION: expanded equation
%%
Inserting the function $B$ one has more explicitly for the leading term 
\be\label{enns}
m(U,V)\simeq {G_{\rm obs}
\over r_0}\Big( 1-{{\cal A}\over V(-\ln (-\kappa_0 V))^{(p+2)}} \Big )
{\dot{A}(U)\over V(-\ln (-\kappa_0 V))^{(p+2)}}
\ee
This is our main result. It confirms our earlier conclusion about the 
damped increase of the mass function within an improved approximation
which takes the back-reaction of the metric into account. 

A priori it might have appeared equally possible to perform
the substitution $G_{\rm obs}\rightarrow G(k)$ directly in the
Einstein equations. The identification $k^2=G(k)m/r^3$ leads to a non-linear equation for 
$k$ that up to $O(k^2/m^2_p)$  is equivalent to $k^2=G_{\rm obs}m/r^3_0$ which was used before.
However it is important to observe that in this case a decreasing $G$ 
leads to the additional effect of lowering the value of the surface gravity at the inner horizon
which, too, damps the increase of the mass function.

It should also be stressed that the above results were obtained by
integrating out only the field modes with momenta between 
$b\sqrt{|\Psi_2|}$ and $m_{\rm p}/a$. While lowering the
IR cutoff even further is difficult from the technical
point of view (the adiabatic approximation is not available any longer)
the monotonicity of the function in eq.(\ref{renorm}) suggests that taking additional modes 
into account will lead to an even stronger damping of the classical increase 
of the mass function. On the other hand, by adding further matter fields
the antiscreening nature of the gravitational interaction could
be destroyed in principle. (In ref. \cite{per} a condition on the number of 
the various species of fields implying $\omega>0$ can be read off.)
We nevertheless believe that the 
qualitative features of our discussion will hold for arbitrary
matter systems with $\omega >0$. In particular for the matter system 
consisting of a massless minimally coupled scalar field
considered in this investigation the quantum-corrected 
geometry is less singular than its classical counterpart.

\section{conclusions}
%%
%% REVISION : see second referee comments.
%%
We have discussed a possible physical mechanism which has the effect of damping 
the classical increase of the mass function behind the 
potential barrier inside a realistic black hole. We believe that this mechanism
is operative already for black holes with masses $M>>m_{\rm p}$ and before
the Cauchy horizon singularity is reached, {\it i.e.} in a regime of
sub-planckian curvatures where it can be calculated reliably. On the basis of the
present investigation we cannot make any claim about the fate of the singularity 
at the CH. However, one can speculate that if the decrease of $G$ continues and that
$G\rightarrow 0$ at the CH, the geometry of the spacetime near the late-time portion 
of the CH is regular with a sub-planckian Weyl curvature. In order to settle
this issue completely a more complete calculation is needed and we hope to 
address this problem in the future.

\vspace{2mm}

It is a pleasure to thank W. Israel for stimulating discussions.
One of us (M.R.) would like to thank the Department of
Physics of Catania University for the cordial hospitality
extended to him while this work was in progress. He is also 
grateful to INFN, Sezione di Catania, for the financial
support which made this visit possible.


\begin{references}
 
\bibitem{pi}E. Poisson and W. Israel, Phys. Rev.
D 41, 1796, (1990).
\bibitem{haifa} {\it Internal structure of black holes 
and spacetime singularities}, volume XIII of the
Annals of the Israel Physical Society, edited by L. M. Burko
and A. Ori (Institute of Physics, Bristol, 1997).
\bibitem{abo}A. Bonanno, S. Droz, W. Israel and S. M. Morsink, 
Proc.R.Soc.Lond. A 450, 553-567, London (1995).
\bibitem{ori}A.Ori, Phys. Rev. Lett. {67}, 1991, 789;
A. Bonanno Phys. Rev. D 53, 7373 (1996).
\bibitem{droz}P. R. Brady, S. Droz, S. M. Morsink,
Phys. Rev. D 58: 080434 (1998).
\bibitem{werlett} R. Balbinot, P. R. Brady, 
W. Israel and E. Poisson, Physics Lett. A 161, 223 (1991). 
\bibitem{israel} W. G. Anderson, P. R. Brady, W. Israel,
S. M. Morsink, Phys. Rev. Lett. 70, 1041 (1993); 
R. Balbinot, E. Poisson, Phys. Rev. Lett. 70, 13 (1993);
\bibitem{corfu} For an introduction see: M. Reuter in
{\it Proceedings of the 5th Hellenic School and Workshop
on Elementary Particle Physics}, Corfu, Greece, 1995 and
hep-th/9602012. 
\bibitem{rw} M. Reuter, C. Wetterich, Nucl. Phys. B 417, 181, 1994;
Nucl. Phys. B 427, 291,1994 .
\bibitem{aboza} A.Bonanno Phys. Rev. D 52, 969, 1995;
A. Bonanno D. Zappal\'a Phys. Rev. D 55, 6135 (1997) 
and hep-ph/9611271.
\bibitem{reuter} M. Reuter, Phys. Rev. D {57}, 971, 1998, 
and hep-th/9605030.
\bibitem{ue}W.Dittrich, M.Reuter, {\it Effective Lagrangians in 
Quantum Electrodynamics, } Springer, Berlin, 1985.
\bibitem{granda}L. N. Granda, Europhys. Lett. {42}, 487, (1998)
\bibitem{per}D. Dou, R. Percacci, hep-th/9707239
\end{references}
\end{document}